# Social capital and small business productivity: The mediating roles of financing and customer relationships

**Christopher Boudreaux**
Florida Atlantic University

**George Clarke**
Texas A&M International University

**Anand Jha**
Wayne State University

**Abstract.** How does an entrepreneur's social capital improve small informal business productivity? Although studies have investigated this relationship, we still know little about the underlying theoretical mechanisms driving these findings. Using a unique Zambian Business Survey of 1,971 entrepreneurs administered by the World Bank, we find an entrepreneur's social capital facilitates small business productivity through the mediating channels of firm financing (i.e., credit from suppliers, credit to customers, loans from friends and family) and customer relationships (i.e., more customers). Our findings, thus, identify specific mechanisms that channel social capital toward an informal business' productivity, which prior studies have overlooked.

**Plain English Summary:** "Small informal businesses are more productive when they build relationships with customers, extend credit to customers, and receive credit from suppliers and friends and family." Using a unique survey of 1,971 Zambian entrepreneurs administered by the World Bank, we find small informal businesses are more productive when they build relationships with customers, extend credit to customers, and receive credit from suppliers and friends and family. Importantly, we observe belonging to a business or social association enables these businesses to extend credit, receive credit, and build relationships. Our findings provide more specific mechanisms that channel social capital toward an informal business' productivity, which prior studies have overlooked. Thus, our main policy recommendation is that entrepreneurs, especially from informal businesses and developing contexts, should focus on networking to foster customer and professional relationships, which can help foster firm productivity.

1. **INTRODUCTION**

Countries grow faster when small businesses are productive and become larger. Small businesses hold more capital, provide more jobs, and are more innovative than large firms (Acs and Audretsch 1988). Small businesses often operate in the shadows in an informal economy that represents more than 50% of a country's gross domestic product (GDP) in emerging economies and about 10–15% GDP in industrialized countries (Schneider and Enste 2013). Given the important role small businesses play in the growth of an economy, academics and policymakers are interested in understanding what makes a small business more productive.

Several studies document a positive association between a firm's social capital—measured by the strength of the owner's network—and the firm's productivity (Fatoki 2011; Bosma et al. 2004; Santarelli and Tran 2013). Although these studies have investigated the relationship between a firm's social capital and productivity, we still know little about the underlying theoretical mechanisms driving these findings in small informal business. Among the few exceptions are, for example, Cooke and Wills (1999) that find that social capital improves performance by enhancing business knowledge and innovation, and reducing transaction costs (Fafchamps and Minten 2002).

The purpose of our study is to identify specific mechanisms that mediate the relationship between social capital and an informal small business productivity, and conduct formal tests for mediating. Because credit constraint is a major obstacle to small business success (Berger and Udell 1995; Cole 1998), we expect credit access to mediate the relationship between social capital and small business productivity. However, all types of credit are not the same. For example, the cost of bank loans, loans from suppliers, and loans from friends and family can be different. The maturity of these debts and the consequences of not paying them on time can also differ. Therefore, the mediating effect might differ based on the type of credit. What is more, other factors, besides access to credit, might mediate social capital's positive effect on productivity. For example, a business-owner with a dense network is also more likely to be trusting (Putnam 2000), and as such, may be better at building good customer relationships and increasing the firm's customer base and therefore increase productivity. While there is a rich discussion on how trade credit can improve small business productivity, a formal test of the mediating effect is lacking. Also lacking is the discussion and evidence that social capital might be associated with more credit to customers and a greater customer base that may lead to higher productivity.

We fill this gap. Specifically, we investigate whether credit from suppliers, credit to customers, loans from family and friends, loans from banks, and the firm's customer base mediates the association between social capital and productivity. We know of no study that examines the mediating role of possible mediators of social capital and small informal businesses' productivity. There are, however, studies showing the effect of the structural dimension of social capital on firm performance is mediated by relational and resource dimensions e.g., Castro and Roldán (2013). We hypothesize entrepreneurs[1] with a high level of social capital will have easier access to credit from their suppliers, banks, and family and friends, and this will, in turn, lead to increases in small business productivity. We also hypothesize that entrepreneurs with more social capital will likely extend more credit to customers, and will have a larger customer base, both of which will mediate the relationship between social capital and small business productivity. We develop our hypotheses using insights from the social capital literature (e.g., Coleman 1988; Adler and Kwon 2002;

---

[1] Although there are many different definitions of entrepreneurship, we consider these Zambian small business owners to be entrepreneurs because they are operating their enterprises to maintain or build their business. Although most of these business owners are not high-tech or growth-oriented, they are all trying to operate in the economy and many have very high entrepreneurial aspirations. This is consistent with calls to embrace entrepreneurial diversity (Welter et al. 2017).

Putnam 2000; Woolcock and Narayan 2000; Lin and Zhang 2009), and the small business financing literature (e.g, Fafchamps and Minten 2002; Biggs and Shah 2006).

To test our hypothesis, we use a unique dataset based on the Zambian Business Survey (ZBS) conducted by the World Bank.[2] It provides us with data on the number of customers and on whether a firm extends credit to its customers, features that make it unique and allow us to examine unique channels by which social capital might improve small business productivity. It is important to study this relationship in a developing economy such as Zambia because social capital matters more when information asymmetries between lenders and borrowers, and between customers and the entrepreneur, are particularly high, which is often the case in developing countries with underdeveloped financial and regulatory systems (Galor and Zeira 1993; Cestone and White 2003). The availability of some unique features in this data and the underdevelopment of the financial market, and poor law enforcement mechanism, make Zambia a good laboratory to examine our hypotheses.

Our results indicate that being a member of an association is associated with a 41 percent increase in productivity, and we also find these entrepreneurs are more likely to provide credit to customers, obtain credit from friends and family, obtain loans from their suppliers, and have a larger customer base. Our findings add to the literature investigating the challenges of operating as a small informal businesses. A long-standing theory in finance suggests a lack of external capital constrains the productivity of small firms. Butters and Lintner (1945), who offer one of the earliest insights into this issue, write: "many small companies—even companies with promising growth opportunities—find it difficult or impossible to raise outside capital on reasonably favorable terms" (page 3). Our study suggests that membership in a business or non-business association positively influences small business productivity, and part of this relationship is because social capital eases credit constraints.

---

[2] One of the co-authors was part of the team that wrote the survey instrument.

We are the first to document that better access to informal credit to the firm is associated with credit to customers and a higher customer base, and greater productivity. Such details on how social capital can improve a small informal business' productivity have been missing in the literature. Hence, we extend the work of Fafchamps and Minten (2002) and Fafchamps (1997). They identify that social capital increases the likelihood of trade credit from suppliers for informal small businesses but do not link it to an extension of credit to customers and productivity.

## 2. THEORETICAL DEVELOPMENT OF HYPOTHESES

### 2.1. Social capital

Following the literature, we define social capital as the networks that facilitate collective action (Granovetter 2018; Fafchamps and Minten 2002). An individual with high social capital is likely to have a bigger network, more friends, and is more likely to have norms that facilitate working together, such as trusting others.[3] Norms and networks are related—participation in an organization develops norms conducive to cooperation such as trust in others, and these norms facilitate participation in networks (Fukuyama 1995). Tsai and Ghoshal (1998) find that "social interaction, a manifestation of the structural dimension of social capital, and trust, a manifestation of its relational dimension, were significantly related," (p. 464). Lyon (2000) summarizes the value of the network in the context of social capital as follows.

---

[3] Fafchamps and Minten (2002) discuss how the norms and network dimension of social capital are related in more detail.

> Networks are the most visible and clearly definable part of social capital and for this reason they have received most attention in studies on social capital. Many analyses, especially those that attempt to quantify social capital, concentrate on formal networks and groups with an assumption that the quality and quantity of associational life can be used as a proxy for social capital (page 676).

**2.2. Social capital and productivity**

We posit an entrepreneur's social capital will positively affect small business productivity because she/he is likely to enjoy greater trust from stakeholders such as lenders and customers and therefore have a lower cost of raising capital. Entrepreneurs with more social capital will also have a lower turnover of employees and good relationships with suppliers.

Empirical evidence supports this view. Using data from Madagascar Fafchamps and Minten (2002) posits that social capital facilitates human interaction, reduces transactions cost with other traders, lenders, and family members, making the firm more productive. They go on to write:

> The strength and robustness of social capital variables stands in sharp contrast with the less robust and partly counterintuitive results obtained with human capital variables such as years of schooling, years of experience as a trader, and the ability to speak more than one language. Although this does not imply that human capital is unimportant, it suggests that social capital might be as, if not more, important for efficiency in economies characterized by high transaction costs and poor market institutions (Fafchamps and Minten 2001).

Leana and Pil (2006) find that internal social capital (relation among teachers) and external social capital (relation between principal and external shareholders) is positively associated with higher performance of students in reading and mathematics. Their logic is that a healthy relationship facilitates greater trust, lower fear of opportunistic behavior, and greater sharing of information.

Following theses lines of thought, we argue that an entrepreneur with high social capital will likely have better relations with stakeholders such as creditors and customers, which will, in turn, improve productivity. In addition to better relationships with stakeholders, social capital enables entrepreneurs to gather useful information which can be used to cut costs and hire and retain good employees leading to higher productivity. So it is not surprising that Stam et al. (2014) conduct a meta-analysis of the association between entrepreneurs and a small firm's performance and find the link to be positive and significant. Based on these studies, we hypothesize the following.

***Hypothesis 1:*** *Entrepreneurs' social capital is positively associated with small business productivity.*

**2.3. The mediating role of credit from suppliers**

There are two types of trade-credits : account receivables—the supply side of trade credit, and account payables—the demand side of trade credit. Both these types of credit are useful for small enterprises.

Small businesses rely heavily on account payables as they cannot issue stocks and bonds in the external market, but have only two sources of external finance: banks and credit from suppliers. According to the U.S. small business administration, about 20 percent of the small businesses in the U.S. do not get financing from banks and rely on credits from suppliers.[4] Berger and Udell (1998) note that a typical small business has an almost equal amount of debt from banks as it has from suppliers. The credit from suppliers is likely to be even higher in developing economies such as Zambia where access to banks is much more difficult, and where there is greater information asymmetry between lenders and borrowers. These countries do not have credit scores for individuals.

---

[4] https://www.sba.gov/advocacy/bank-credit-trade-credit-or-no-credit-evidence-surveys-small-business-finances

The importance of credit from suppliers is well established in the small business literature. For example, Wilner (2000) finds that access to credit from suppliers is particularly useful when the firm faces an unexpected financial constraint. Cunat (2007) echoes the same findings: suppliers of a firm can be a source of mitigating financial constraints when the firm faces temporal liquidity shocks threatening their survival. In a sample consisting of about 200,000 small and medium enterprises spanning across 13 European countries, McGuinness et al. (2018) find that higher levels of account payables during financially difficult times is associated with higher survival rates. Ogawa et al. (2013) find that credit from a supplier is associated with higher profitability for small businesses.

There are also theoretical underpinnings for linking access to credit from a supplier to a firm's productivity. Agostino and Trivieri (2019) argue that it helps "to smooth out the production process and avoid inventory shortages and the associated interruptions or inefficiencies in production," (page 577). It makes it easier to verify the product's quality before payment, making it easier to return defective products (Long et al. 1993). Trade credit from suppliers also improves efficiency because it makes possible agreements that stagger the payment and deliveries over time, making it easier for the firm to handle their working capital (Ferris 1981; Schwartz 1974; Fisman 2001).

Possibly, firms that have greater access to supplier's credit may be in a better position to extend credit to customers. Because they are similar in their maturity, access to credit from a supplier might be the most effective type of financial resources useful in extending credit to customers. Deloof and La Rocca (2015) who examine small businesses in Italy find that small businesses that have better access to finance provide more credit to customers.

We posit that the high social capital of the entrepreneur will increase the chance of an owner obtaining supplies on credit. Credit from suppliers does not depend upon how much collateral the small business has, but rather on trust and reputation (Fafchamps 1997), and suppliers will have higher trust in clients with high social capital. They will, rightly, expect ethical behavior from their client, and expect repayment on time. Research shows that high social capital individuals are less likely to commit a crime (Buonanno et al. 2009). Experimental studies also show that high social capital individuals are more likely to repay their debt (Karlan 2005). Because suppliers are in closer relations with their clients, they are in a better position to factor in soft information such as trust.

Therefore, we hypothesize the following:

***Hypothesis 2a:*** *Credit from suppliers will mediate the association between social capital and small business productivity through the number of customers.*

***Hypothesis 2b:*** *Credit from suppliers will mediate the association between social capital and small business productivity through credit to customers and the number of customers*

**2.4. The mediating role of credit to customers**

While better access to credit from suppliers helps the financing needs of small businesses, extending credit to customers also offers many advantages. The value of extending credit is summarized in the following excerpt from Wilson and Summers (2002).

> Trade credit is also an important tool for relationship building and management. Thus, on the supply side trade credit [account receivables] can be a multi-faceted and important strategic or competitive tool that plays a role in capturing new business, in building supplier customer relationships (developing an implicit equity stake in the customer), in signalling product quality, `reputation' and financial health, and in price competition and price discrimination (page 317).

Building a customer base by providing account receivables to the right customers is vital for small business success. Martínez-Sola et al. (2014) examine over 10,000 manufacturing SMEs from Spain and find a higher proportion of account receivables is associated with higher profitability. Moreover, they find providing credit to customers is particularly useful in situations with variable demand as it can smooth demand, lower operating costs, and enhance productivity.

We posit that high social capital will be associated with a greater extension of credit to customers because an owner who is more trusting of others is more willing to be vulnerable to his customer defaulting. Research shows that the extent to which you trust others affects whether you part with your cash and invest. For example, Guiso et al. (2004) find that high social capital individuals hold less cash, use more checks, and participate more in the financial market. Moreover, when the owner has a high level of social capital, she will likely have better relationships with customers, which can reduce default rates for two reasons. With better relationships, comes more accurate information, which she/he can use to extend credit to only worthy customers. Second, better relations increases the customers' cost of defaulting. Field experiments suggest that clients are also less likely to default when there is higher personal trust (Cassar et al. 2007).

In developing economies, where clients are credit constrained, inability to provide credit to customers may mean losing customers. Most customers would prefer the flexibility of obtaining goods on credit as it makes their financial planning easier. Schwartz (1974) considers it an "integral part of a firm's pricing policy," (page 644). Schwartz and Whitcomb (1977) theorize that firms can use it to disguise price discrimination by charging clients that delay payment at a slightly higher price. Wilson and Summers (2002) note extending credit to customers may also help build customer loyalty. It is also a tool to signal to customers that they are financially sound. The following quote is illustrative of theoretical reasons why providing customers goods on credit might enhance productivity:

> Small firms which are startups or have aims for growth also face problems of reputation when entering new markets. They may need to use trade credit as a signal of reputation and commitment, and as a marketing tool. This latter influence has some echo in Petersen and Rajan's (1997) finding that there is a `greater extension of credit by firms with negative income and negative sales growth', where they suggest trade credit is used as a signal of financial health and to boost sales (p. 318).

Therefore, we hypothesize the following.

***Hypothesis 3:*** *Credit to customers will mediate the association between social capital and small business productivity through a larger customer base*

**2.5. The mediating role of loans from friends and family**

S. Lee and Persson (2016) note that several million small businesses from forty-two countries raised about 600 billion from informal investors, and some rely exclusively on this type of finance. They also posit that most of the informal financing comes from family and friends. Though many entrepreneurs may not prefer to borrow from family and friends fearing strain in social ties, in poor economies this may be the only option, since getting a loan is difficult. Moreover, borrowing costs are lower when borrowing from friends and family, partly because the classic problem of moral hazard and adverse selection is lower (Banerjee et al. 1994; Stiglitz 1990). An entrepreneur with high social capital, because of her/his reputation and participation in various networks will be more likely to obtain loans from family and friends, increasing the size of the customer base and enhancing productivity. Because credit constraints are a serious concern for small businesses in poor developing economies (Fafchamps 1997), financial help from family and friends might provide funds to buy machinery, reach new markets, and expand when opportunities arise. Hence, we hypothesize the following:

> ***Hypothesis 4:*** *Loans from friends and family will mediate the association between social capital and small business productivity through the customer base*

### 2.6. The mediating role of loans from banks

The literature on small business lending, social capital, and small business performance suggests access to bank loans might mediate the association between social capital and small business productivity. Pham and Talavera (2018) examine micro, small, and medium-sized firms in Vietnam and find that owners who have bigger networks are likely to have better access to loans from banks. They argue that participation in social networks increases access to loans because it allows banks to gather more information (Le and Nguyen 2009). Talavera et al. (2012) find that small business owners who are more altruistic, measured by whether they contributed to a charity in the past, are more likely to get their loan application approved. Their argument is similar—individuals who contribute to charity are embedded in networks, reducing the transaction cost of private information sharing among lenders and borrowers (Uzzi 1999; Boot 2000). Hernández-Cánovas and Martínez-Solano (2010) analyze the relationship between banks and small businesses in Europe and find trust between the firm and bank improves access to finance and borrowing costs. Hence, we make the following hypothesis:

> ***Hypothesis 5:*** *Loans from banks will mediate the association between social capital and small business productivity.*

### 2.7. The mediating role of customer base

Customer service is important for firms desiring to increase the number of customers. Research sponsored by Zendesk, a customer service company, found 58 percent of respondents stopped buying from a company after they experienced bad service, and 52 percent told others not to buy from the firm.[5]

Entrepreneurs with more social capital are likely to be cooperative and trusting of others, which is likely to build better relationships with customers, provide better service, and therefore have a bigger customer base. Owners of small businesses who are more trusting may offer better services and hence increase their customer base. Merlo et al. (2006) examines customer service in retail firms in the U.S. and finds, unsurprisingly, that a trusting culture is associated with greater customer satisfaction.

There are other ways in which social capital increases customer bases. Yli-Renko et al. (2001) argue, and provide empirical support for their idea that the social network dimension of social capital helps owners acquire knowledge from various sources including the customers, and that this knowledge can be exploited for comparative advantage and increasing sales. For example, they may be better at distinguishing between customers who are worthy of extending credit. Fafchamps and Minten (2002) examine the returns to social networks among agricultural traders in Madagascar and document how traders with larger networks, measured by the number of other traders they know, have larger sales performance. They argue this is due to a lower transaction cost of finding other traders. Entrepreneurs who have a bigger network may identify potential customers, or acquire knowledge that could help differentiate their products to attract more customers, and this in turn will increase small business productivity. Moreover, because they are relatively less credit constrained themselves they may be at a better position to extend credit to customers. Thus, we propose the following hypothesis:

> ***Hypothesis 6:*** *The size of the customer base will mediate the relationship between social capital and productivity.*

---

[5] https://www.zendesk.com/resources/customer-service-and-lifetime-customer-value/

## 3. DATA and EMPIRICAL MODEL

This paper uses data from the nationally representative Zambia Business Survey (ZBS). The survey, which was conducted in late 2008 by Finmark and the World Bank, includes 4,801 small businesses with 50 or fewer workers. However, because we exclude agricultural firms, we are left with approximately 2,000 businesses in our sample.

### 3.1. Data and Summary Statistics

*3.1.1 Sample*

The ZBS covers commercial firms that produce goods or services that are sold to people or firms outside the owner's household. The survey firm (Steadman Research Services) selected firms using area sampling.[6] First, the survey firm randomly selected 320 enumeration areas (EAs) from a stratified list based on the 2000 census. Once the survey firm had selected the EAs, they listed all houses and other buildings in the area and checked whether the buildings contained people who owned and ran their own businesses. They used this to make a list from which they randomly selected people. The sample, therefore, includes small, home-based firms as well as formal businesses.

---

[6] See Clarke et al. (2010) for a more detailed description of the survey.

The study focuses on non-agricultural firms.[7] We exclude farms for two reasons. First, many small farms are not commercial firms—they are subsistence farms that sell excess production. Second, it is difficult to measure how much these farms produce. To work out how much the farm produces, farmers must estimate how much their family consumes and how much it is worth. These imprecise estimates mean it is difficult to estimate the productivity of subsistence farms.[8]

The firms were mostly small shops—about 75 percent of the sample. The remaining firms were in manufacturing (10 percent of firms), services (15 percent) and other areas such as mining, health, and electricity (1 percent). Although the sampling frame included firms with up to 50 workers, few were this large. Firms had an average of 1.8 paid workers (including the owner) and 1.7 unpaid workers (often family members). The sample, however, was skewed. About 78 percent of firms had no paid workers except the owner, and 57 percent had no paid or unpaid workers except the owner. Only 10 percent of firms had more than 5 workers (paid or unpaid), and only 5 percent had more than 10 workers (paid or unpaid).

Few firms use sophisticated production methods. Only 18 percent had electricity, only 14 percent had water from a public source, and only 2 percent used a fixed line phone. About 43 percent of firms had or used a calculator, about 5 percent had or used a car, and only 2 percent had or used factory machinery. In addition, only about 14 percent of firms used a business bank account and only 6 percent of firm owners used a personal bank account.

---

[7] The data in this section refers to the non-agricultural firms in the survey that the empirical work focuses on.

[8] Consistent with the idea that the estimates are imprecise, most estimated that self-consumption was a round number. Almost half said either 20, 30 or 40 percent of output.

Most were also informal—only 15 percent of owners said they had registered their firm with any government agency. Further, most had only registered with their local government. Only 5 percent had done so with the national tax authority (Zambia Revenue Authority) and only 8 percent with the company registrar (Patents and Companies Registration Office).[9]

The ZBS is, therefore, different from other Zambian surveys such as the Regional Program for Enterprise Development (RPED) surveys and the World Bank's Enterprise Surveys (WBES) that have focused on larger formal firms. The 2013 WBES, for example, explicitly excluded informal enterprises (World-Bank 2009).[10] As a result, the firms were far larger than those in the ZBS—the mean and median number of employees were 54 and 15. Only 10 percent of ZBS firms had more than 5 employees—the minimum size for the WBES. The earlier RPED survey also focused on larger enterprises; the mean and median number of employees were 85 and 23 respectively (Van Biesebroeck 2005).[11] Although the RPED survey included a few informal enterprises, the survey did not sample them systematically.[12]

---

[9] Moreover, because this information is self-reported it probably overestimates registration.

[10] Moreover, the sample was based on a sampling frame given by the Zambia Central Statistical Office, which implicitly excludes informal firms as well.

[11] Biggs and Shah (2006), for example, use this data.

[12] Van Biesebroeck (2005) reports that the 'selection of informal firms was generally left to the interviewers" (p. 549).

*3.1.2. The Zambian context*

Zambia is a low-income country categorized by high rates of self-employment and high rates of entrepreneurial aspirations.[13] According to a Global Entrepreneurship Monitor 2013 Global Report, in 2013, 40 percent of individuals were involved in total early-stage entrepreneurial activity (TEA), but less than half that number (16.6 percent) actually established a business. This suggests much of the entrepreneurship in Zambia is informal. Although there is some variation over time, the long-run trends are relatively stable. According to the OEC,[14] two of Zambia's largest trading partners are South Africa and the Democratic Republic of Congo. However, Zambia has an even larger trading relationship with Switzerland and China. As a result, there is significant intraregional trade, but the trading relationships do extend beyond the continent. Zambia's largest export is copper, both raw and refined, but Zambia also exports raw tobacco and postage stamps. According to the Global Entrepreneurship Index 2018 report,[15] Zambia also ranks 102 out of 137 countries, which places it in the bottom quartile of the rankings. Zambia's GDP is about 70 billion, and its GDP per capita in 2017 dollars is around $4000 (in terms of PPP). Based on 2015 data, 54.4 percent are below poverty[16]. Further, according to most estimates[17], Zambia's informal Sector employs about 90 percent of its labor force. For these reasons, Zambia is a good laboratory to study this relationship because social capital matters more when information asymmetries between lenders and borrowers, and between customers and the entrepreneur, are particularly high, which

---

[13] https://www.gemconsortium.org/economy-profiles/zambia

[14] https://oec.world/en/profile/country/zmb/

[15] https://thegedi.org/downloads/

[16] https://www.cia.gov/library/publications/the-world-factbook/geos/za.html

[17] https://www.theigc.org/wp-content/uploads/2012/06/Kedia-Shah-2012-Working-Paper.pdf

is often the case in developing countries with underdeveloped financial and regulatory systems (Galor and Zeira 1993; Cestone and White 2003).

*3.1.3. Measuring social capital*

We consider the entrepreneur to have high social capital if she/he is a member of an association. Our operationalization of social capital is consistent with the literature. Benson and Clay (2004), for example, use the frequency of church attendance and marital status as a measure of social capital. Many other studies use the size of the network as a measure of social capital (e.g., Fafchamps and Minten 2002; Granovetter 2018; Kreiser et al. 2013). In Stam et al. (2014) who conduct a meta-analysis of 61 independent samples to examine the effect of social capital on performance, the measure for social capital is the size and intensity of networks.

The variables that most interest us measure the owners' social capital. We measure social capital using a dummy, *Any association,* coded 1 if the owner belongs to any assocciation and 0 otherwise. Associations can be a business association or a non-business association.[18] Non-business associations include churches, religious groups, political parties, women's or men's groups, social clubs, and sports clubs. Owners who belong to these might use them to meet potential customers or employees or people who can help them with commercial or technical issues. Owners who belong to these groups might also meet people who can help them get trade or bank credit.

Many firm owners belonged to social groups—about 67 percent of owners. Fewer belonged to business associations—only about 5 percent. Owners who belonged to business associations were more likely to also belong to a social group than other owners—86 percent compared with 66 percent.

---

[18] In additional robustness checks, we separate this variable into business and non-business associations. Both measures are dummy coded (1=yes; 0=no).

Table 1 reports the summary statistics. About 6.6 percent of the firms in our sample obtain credit from suppliers, and about 31.2 percent provide credit to customers. Less than three percent get a loan from banks, and about 5.6 percent get loans from family and friends.

-----------------------------------------------
INSERT TABLE 1 ABOUT HERE
-----------------------------------------------

Firms with owners who belonged to business associations and social groups were 3.2 percent more likely to have had loans from friends and family than firms with owners that did not belong. The difference is statistically significant at a 5 percent significance level. Firms with owners who belonged to business associations and social groups were also 6.8 percent more likely to provide credit to customers, 2.5 percent more likely to receive credit from suppliers, and have 5.9 more customers on average, when compared to firms with owners that did not belong. These differences are statistically significant at a 5 percent significance level, but there was no statistical difference between these firms for bank loans.

Owners who belonged to a business association or social group were also more likely to provide customers with credit and to receive credit from suppliers. About 36 percent of members provided credit to customers and 8 percent of members received credit from suppliers. In comparison, only 31 percent of non-members gave credit and only 5 percent received credit. The differences are statistically significant. In contrast, firms with owners that were members did not have more customers than firms whose owners were not. The median firm in both groups reported between 11 and 50 customers in a month.[19]

### 3.1.4. Measuring productivity

We measure productivity as the ratio of sales to the number of workers. This measure is a simple, elegant, and legitimate measure of productivity. Bloom et al. (2010) use the ratio of sales to the number of workers to understand why productivity is lower among manufacturing firms in developing economies. Because of its simplicity and understandability, it is a common measure of productivity (Mahmood 2008). In additional robustness checks (Table A2 in the online appendix), we show that different assumptions regarding how to treat part-time and unpaid workers does not affect the results. Our results are robust when excluding firms with any part-time workers or any unpaid workers as well as counting part-time workers as half a worker.

### 3.1.5. Control Variables

We include several variables to control for characteristics of the owner, the firm, and the community where the firm operates. We control for the age, experience and education of the owner. *Age of entrepreneur* is measured as the natural log of the business owner's age. Firms with older owners might perform better if the age of the owner is a reasonable proxy for experience. Previous studies using enterprise level data for Sub-Saharan Africa have found that firms perform better when the owner is better educated (Biggs et al. 1998; Ramachandran and Shah 1999). Therefore, we include measures of the owner's education. *University education* is coded 1 if the owner has a university education and 0 otherwise. *Vocational education* is coded 1 if the owner has a vocational education and 0 otherwise. *Secondary education* is 1 if the owner has a secondary education and 0 otherwise. The omitted category is primary education or less. *Has bank account* is coded 1 if the owner has a personal bank account and 0 otherwise. Possessing a personal bank account is a signal that the manager is financially sophisticated, which should hopefully translate into better firm management. We include the age of the firm as another control. *Firm age* is measured as the natural logarithm of the age of the firm.

---

[19] Owners responded with ranges rather than exact numbers. The ranges were: 0 customers; 1-5 customers; 6-10 customers; 11-50 customers; 51-100 customers; 101-500 customers; 501-1000 customers; and over 1000 customers.

For our mediating analysis, we also include firm financing variables and information on the customer base. *Number of customers* is measured as the natural logarithm of the firm's number of customers. *Obtains credit from suppliers* is coded 1 if received credit from suppliers and 0 otherwise. *Provides credit to customers* is coded 1 if provides credit to customers and 0 otherwise. *Loan from bank* is coded 1 if the firm received a loan from a bank and 0 otherwise. *Loans from friends and family* is coded 1 if the firm received a loan from a friend or family member and 0 otherwise.

In addition to owner and organization attributes, it is important to capture regional variation that might influence small business performance. We include several regional-level variables to control for agglomeration effects. *Firm is in urban area of district* is coded 1 if the region is categorized as either urban or peri-urban and 0 otherwise. This designation is based on the classification in the 2000 Census, which was used for sampling. This is used as a proxy for the potential presence of economies associated with agglomeration e.g., Audretsch et al. (2015). *Population density* is measured as the population per square kilometer. The population density is included as an additional measure of the agglomeration effects. When firms are close to customers, workers and suppliers, they might find it easier to share knowledge or pool capital, intermediate inputs and labor. Agglomeration could also improve matching between firms and their customers, suppliers and workers. *Illiteracy rate* is the measured rate of illiteracy. It is included as it might potentially affect knowledge sharing. Lastly, we included several dummy variables for information on sectors—retail, manufacturing, services, and other.

We observe several relationships in the data according to the correlation matrix in Table 2. We observe that our measure of social capital, *Any association*, is positively correlated with labor productivity. However, the correlation appears to be driven by the non-business association rather than membership in a business association. Firms that have more customers also appear to be more productive and all measures of financing—except for bank loans—are positively correlated with labor productivity. We also observe a positive correlation between business associations and non-business associations but it is small in magnitude (r=0.069). This suggests that those who belong to one type of association are more likely to belong to the other, but there is also variation in the data.

-----------------------------------------------
INSERT TABLE 2 ABOUT HERE
-----------------------------------------------

The empirical model can be formally stated as:

$$DV = \beta_0 + \beta_1 SC + \sum_{i=1}^{7} \gamma_i FC_i + \sum_{j=1}^{2} \delta_j DC_j + \mu_k + \varepsilon \qquad (1)$$

Where DV denotes our dependent variable (provides credit to customers, gets credit from suppliers, loan from friends/ family, loan from bank, number of customers (log), labor productivity), SC denotes our measure of social capital, FC is a vector of firm characteristics (owner has university education, owner has vocational education, owner has secondary education, firm age (log), firm is in urban area, age of entrepreneur, firm has bank account), DC is a vector of district characteristics (population density, illiteracy rate), $\mu$ is a set of industry fixed effects, and $\varepsilon$ is the stochastic error term. Depending on the dependent variable, the regression model is either estimated by Probit regression, interval regression, or ordinary least squares (OLS) regression. For models with binary dependent variables (i.e., provides credit to customers, gets credit from suppliers, receives loan from friends/ family, receives loan from bank), we estimate the model using Probit regression. For the model with the number of customers (log), we estimate the model using interval regression, and for labor productivity we use OLS regression.

## 4. RESULTS and DISCUSSION

### 4.1. Regression Results

We begin the empirical analysis in Table 3, which presents the main findings with respect to our financing variables—credit to customers (Model 1), credit from suppliers (Model 2), loans from friends and family (Model 3), and loans from banks (Model 4)—as well as the number of customers (Model 5) and labor productivity (Model 6). Except for bank loans, social capital has a positive and statistically significant coefficient in all models. The findings in Table 2 indicate founders who belong to associations are more likely to: provide credit to customers ($\beta=0.196$; $p<0.01$), receive credit from suppliers ($\beta=0.203$ ; $p<0.05$), receive loans from friends and family ($\beta=0.278$; $p<0.01$), have more customers ($\beta=0.155$; $p<0.05$), and have higher labor productivity ($\beta=0.350$; $p<0.01$). In contrast, belonging to an association does not affect one's chances to receive a bank loan ($\beta= -0.05$; $p>0.10$). Because we include several control variables in our reduced form analysis, we conclude that these relationships are highly robust. For example, these relationships are robust to the inclusion of several firm characteristics including the entrepreneur's level of

education—university, vocation, or secondary—as well as the age of the firm, age of the entrepreneur, and whether or not the firm is located in an urban area. These findings are also robust to the inclusion of district characteristics—population density and the illiteracy rate—as well as industry fixed effects.

More than merely statistical significance, we also observe large effect sizes for most of these relationships. The results (summarized in Table A1 in the online appendix) suggest that, compared to those who do not belong to an association, those who do belong to an association are 6.8 percent more likely to provide credit to customers, 2.5 percent more likely to receive credit from suppliers, 3.2 percent more likely to receive loans from friends and family, and would have 5.9 more customers on average.

---------------------------------------------
INSERT TABLE 3 ABOUT HERE
---------------------------------------------

**4.2. SEM Analysis**

Our results have established that social capital affects the credit from suppliers, credit to customers, loans from friends and family, the number of customers, and labor productivity. We now turn our attention to the main contribution of our study—to investigate the possible mediating channels of firm financing and the customer base.

To test our mediating hypotheses, we use SEM analysis to investigate the multiple channels through which social capital can influence labor productivity. SEM is useful because it allows us to investigate both the direct effects as well as the indirect effects in a model (Hoyle 1995), and it allows for the errors in the structural equations to be correlated with each other (Shaver 2005). In this analysis, we examine a direct path from social capital to labor productivity as well as an indirect path via financing channels and the firm's customer base. SEM allows us to examine whether and to what extent these indirect paths mediate the relationship between social capital and labor productivity. In the SEM analysis, we include, but do not report, all control variables from Table 3. Also, to ease interpretation, we standardized all coefficients in the SEM analysis (i.e., Figure 1 and Table 4).

Our results provide strong support for many of our hypothesized relationships. First, we observe a direct relationship between social capital and labor productivity ($\beta= 0.271$; $p<0.01$). Second, we observe an indirect relationship that runs from social capital to firm financing, which in turn affects the number of customers and ultimately labor productivity. More specifically, social capital positively affects credit from suppliers ($\beta= 0.408$; $p<0.10$), credit to customers ($\beta= 0.273$; $p<0.05$), and loans from friends and family ($\beta= 0.588$; $p<0.01$). Consistent with our results in Table 2, we do not observe any relationship between social capital and bank loans ($\beta= -0.094$; $p>0.10$). Next, we find that credit from suppliers has a positive effect on credit provided to customers (($\beta= 1.72$; $p<0.01$). In turn, we observe that the credit provided to customers ($\beta= 0.239$; $p<0.01$) and loans from friends and family ($\beta= 0.255$; $p<0.10$) both positively influence the number of customers. Lastly, the number of customers positively influences labor productivity ($\beta= 0.203$; $p<0.01$).

In the SEM model, we have to include the number of customers as a continuous variable rather than range (i.e., as an interval). In the main results, we use the top of the range as the variable. In practice, however, the results are not highly sensitive to other points in the range. In particular, the results are similar in terms of size and significance of the results if we use the bottom of the range instead of the top.

-----------------------------------------------
INSERT FIGURE 1 ABOUT HERE
-----------------------------------------------

Table 4 summarizes the direct and indirect channels through which social capital influences firm labor productivity. The largest indirect effect[20] runs (β=0.034) from social capital to credit from suppliers, which in turn, influences the credit given to customers, which then influences the number of customers and ultimately labor productivity. This indirect channel accounts for 9.4 percent (.034/.362) of the total effect. We also observe an indirect effect running from social capital to credit to customers, which then influences the number of customers and ultimately labor productivity. This indirect channel accounts for four percent (.013/.362) of the total effect. Next, we observe an indirect effect running from social capital to loans from friends and family, which in turn, influences the number of customers and ultimately labor productivity. This indirect channel accounts for seven percent (.027/.362) of the total effect.

We do not find evidence that loans from banks mediate the association between social capital and productivity. One explanation is that, while banks rely on soft information, they also need some level of hard information. For example, they are likely to make sure the business is registered with the government. As we noted, only 15 percent of owners were registered with any government agencies. It is not surpring then that only less than four percent of owners have a bank loan. It is possible that those who are members of a business organization are more likely to be registered. Indeed, we find in our supplemental analysis (available in the online appendix) when owners are members of a business organization, they are significantly more likely to receive a loan.

---

[20] Indirect effects are calculated by multiplying all coefficients along the path of analysis. For example, we calculate the indirect effect of social capital on labor productivity *through the channel of loans from family and friends* as the following: $(.588 \times .225 \times .203) = 0.027$.

Our results also suggest that the mediating effect of account payables comes mainly from increasing the likelihood of extending credit to consumers. To calculate the proportion mediated, we follow Alwin and Hauser (1975) and the literature on "inconsistent" mediation MacKinnon et al. (2007)—i.e., when paths have opposite signs. This literature suggests researchers take the absolute value of all path coefficients prior to summation.[21] Overall, we find that 25.14 percent of the relationship between social capital and labor productivity is mediated by these financing and customer channels.[22]

---
INSERT TABLE 4 ABOUT HERE
---

**4.3. Robustness checks**

We conducted a variety of robustness tests to check the sensitivity of the association between social capital and labor productivity. We present those results in an online appendix but briefly summarize them here. Our first check was to assess whether treating the relationship between social capital and productivity as endogenous affects our results. We estimated an Instrumental Variable (IV) model using two instruments for our endogenous variable, social capital—the percentage of the district's business founders who speak the same main language as the founder, and the percentage of the district's business founders who belong to business and nonbusiness groups. These instruments should satisfy the exclusion restriction. For example, founders who speak the same language as other business owners in the same district might be more likely to join such organizations than are other owners. Similarly, founders might be more likely to join groups when other owners in the same district belong to them. We do not expect that owners' group membership or other owners' ethnicity will directly affect the firm's productivity. The founder's ethnicity might, however, affect productivity because of ethnic business networks. The results using the IV approach are quantitatively similar. The relationship between social capital and labor productivity is also robust to alternative modeling methods including estimating a least absolute deviations (LAD) model rather than linear regression. The results in all models are very similar. We estimate our model using paid and unpaid workers together in one model as well as separately. The results are similar either way. Lastly, in additional robustness checks, we include ethnic dummies for the founder's ethnicity, following Biggs and Shah (2006). We also include additional sector, industry, and district dummies as well as different categories of firms including informal firms only, microenterprises, retail only, and informal enterprises with no workers. In all models, we observe a positive and statistically significant effect of social capital on labor productivity. While these sensitivity checks say little about the mediating relationships, they do suggest our findings that social capital positively influences small business productivity are not sensitive to the modeling methods, sample, and measure of productivity.

---

[21] Indirect effect = 0.091 = 0.034 + 0.013 + |-0.013| + |-0.004| + 0.027; Total effect = 0.362 = 0.271 + 0.091.
[22] Proportion mediated = indirect effect / total effect. i.e., 25.14% = 0.091/0.362.

## 5. CONCLUSION

In this study, we investigate the underlying theoretiecal mechanisms that mediate the relationship between an entrepreneur's social capital and small business productivity. Using the owner's membership in an association as a measure for social capital, we find that an entrepreneur's social capital improves the productivity of these enterprises in several ways. First, social capital increases access to credit from suppliers and loans from family and friends. Second, social capital increases the likelihood of providing credit to customers and the size of the firm's customer base. Third, we find that entrepreneur's with more social capital are more likely to receive credit from suppliers, which in turn, makes them more likely to extend credit to customers. Social capital, thus, can facilitate the acquisition of customers and ultimately productivity by operating through the channels of firm financing. Lastly, we do not find evidence that membership in associations makes one more likely to receive a bank loan.

### 5.1. Managerial and Policy Implications

Our study has several managerial and policy implications. One managerial implication of our study is that entrepreneurs might want to become actively involved in various organizations and associations. Doing so might build social capital, increase access to financing, and attract a larger customer base, all of which might increase small business productivity. Related to this, these social capital effects work in tandem. Investing in one's social capital might benefit the entrepreneur because he or she now has greater access to financing through loans from friends and family and trade credit. This credit, in turn, could be extended to more customers to expand the firm's customer base. A policy implication is that local governments and organizations might look to encourage greater networking amongst small business owners and local entrepreneurs. Our results suggest that encouraging people to participate in associations can increase small business productivity by allowing better access to credit, and building better customer relations. Social capital is generally low in many developing countries. Even developed countries such as the United States (Putnam 2000) have seen a decline in the last several decades. Therefore, local policymakers must think of new ways to attract individuals from similar and also different backgrounds.

### 5.2. Contributions to the firm financing literature

Our study's important takeaway is how social capital improves informal small business productivity in developing countries by facilitating everyday financing (i.e., friends, family, and suppliers) but not funding formal (i.e., bank loans). The literature has alluded to this possibility as institutions are weak in these countries, and informal business owners often lack the sophistication necessary to obtain loans (Qian and Strahan 2007). But we do not know of any formal test that links social capital, informal financial financing, and productivity.

We also complement the work by Biggs and Shah (2006) who find enterprises owned by people of Asian or European origin are more productive. They use a sample that mostly comprises large formal firms in sub-Saharan Africa. The authors argue these firms have better access to supplier credit, presumably from their networks, and start as significantly larger enterprises. Our study extends their work by documenting that one does not need to be a member of a specific ethnic group to perform better: membership of any type of association also confers advantages to people regardless of their ethnicity. In our sample, only about 3 percent of the firms are owned by people who belong to Asian and European ethnic groups, and our results are robust when we control for this and when we remove such firms from our sample.

### 5.3. Contributions to the entrepreneurship literature

We contribute to the literature on social networks and entrepreneurship. This literature documents a strong link between an entrepreneur's social network and the identification of new opportunities (Ellis and Pecotich 2001; Elfring and Hulsink 2003; Arenius and De Clercq 2005; Deller et al. 2018), access to foreign markets (Ellis 2000; Zhou et al. 2007), firm entry (Bastié et al. 2013; Kim et al. 2006), the accumulation of knowledge and knowledge spillovers (Sapienza et al. 2005; Hayter 2013), and new venture performance (Brüderl and Preisendörfer 1998; Cooke and Wills 1999; Westlund and Bolton 2003; Santarelli and Tran 2013). Our study complements these studies by showing an entrepreneur's social network facilitates small business productivity, through the mediating channels of obtaining credit from suppliers, receiving loans from friends and family, increasing the likelihood of extending credit to customers, and increasing the customer base.

More broadly, our study contributes to the social capital literature related to small business by documenting the channels by which social capital boosts business development and entrepreneurship (Adler and Kwon 2002; Bosma et al. 2004; Boudreaux et al. 2018; H. S. Lee 2017; Ramachandran and Shah 1999; Ramcharran 2017; Motta 2020; Owalla et al. 2019). While this literature documents that social capital encourages firm performance, we dig deeper into this relationship to uncover the specific mechanisms underlying these relationships.

### 5.4. Limitations and suggestions for future research

As with any other study, our study faces a few limitations that should be addressed by future research. One limitation is its external validity. It is important to note that our findings rely on data from entrepreneurs in Zambia and is probably not unique to Zambia and extends to other developing economies, but this is left to future research.

. Our study also raises the possibility of testing other mediating variables. The mediating variables we examine explain about 25 percent of social capital's effect on labor productivity. It is possible that other mediating variables, such as employee turnover, employee selection and retention also play a role. Analyzing these additional mediating channels is beyond the scope of our study because of data constraints, but would add to our understanding of how social capital improves a small business's productivity

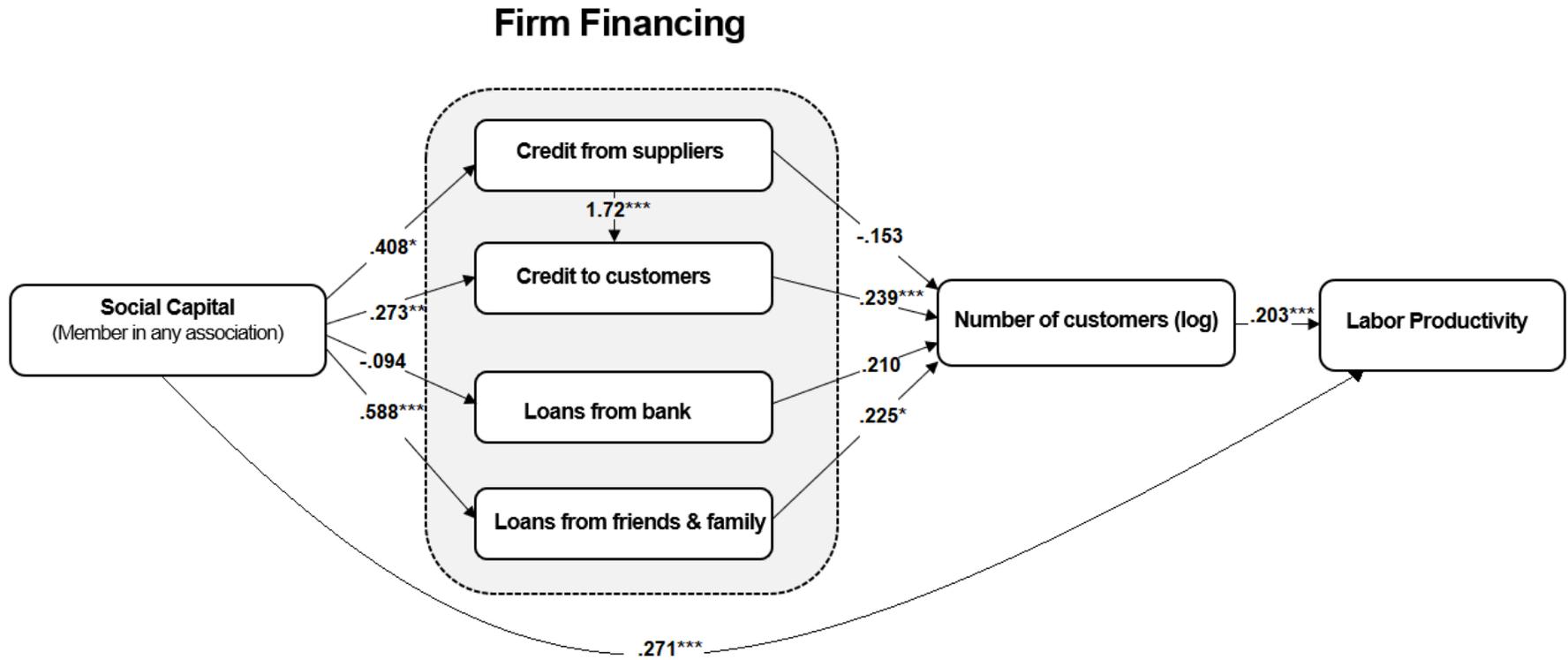

**Figure 1.** SEM Model.
*Note*. The model includes all basic controls from Table 2 and was estimated using Stata's gsem command using maximum likelihood $N= 1,971$.
***$p < .01$. **$p < .05$. *$p < .10$

**Table 1.** Sample means for dependent and independent variables

|  | Mean | Std. Dev. | Min | Max |
|---|---|---|---|---|
| **Productivity** | | | | |
| Labor Productivity (log) | 14.6 | 1.8 | 3.5 | 23.6 |
| **Social Capital** | | | | |
| Any association (dummy) | 0.680 | 0.466 | 0.000 | 1.000 |
| Non-business association (dummy) | 0.670 | 0.470 | 0.000 | 1.000 |
| Business association (dummy) | 0.051 | 0.219 | 0.000 | 1.000 |
| **Mediators** | | | | |
| Number of customers (log) | 4.1 | 1.2 | 1.6 | 6.9 |
| Obtains credit from suppliers (dummy) | 0.066 | 0.249 | 0.000 | 1.000 |
| Provides credit to customers (dummy) | 0.312 | 0.463 | 0.000 | 1.000 |
| Loan from bank (dummy) | 0.024 | 0.153 | 0.000 | 1.000 |
| Loan from friends/family (dummy) | 0.056 | 0.231 | 0.000 | 1.000 |
| **Firm Characteristics** | | | | |
| Founder has university education (dummy) | 0.069 | 0.254 | 0.000 | 1.000 |
| Founder has vocational education (dummy) | 0.064 | 0.244 | 0.000 | 1.000 |
| Founder has secondary education (dummy) | 0.513 | 0.500 | 0.000 | 1.000 |
| Firm Age (log) | 1.6 | 0.9 | 0.0 | 4.6 |
| Firm is in urban area of district (dummy) | 0.457 | 0.498 | 0.000 | 1.000 |
| Age of entrepreneur (log) | 3.6 | 0.3 | 2.8 | 4.5 |
| Has Bank Account (dummy) | 0.056 | 0.231 | 0.000 | 1.000 |
| **District Characteristics** | | | | |
| Population density (log) | 4.3 | 2.2 | 1.1 | 8.5 |
| Illiteracy rate (percent of population) | 30.4 | 14.0 | 10.7 | 66.4 |
| **Sectors** | | | | |
| Retail Trade (dummy) | 0.745 | 0.436 | 0.000 | 1.000 |
| Manufacturing (dummy) | 0.096 | 0.294 | 0.000 | 1.000 |
| Other Services | 0.151 | 0.358 | 0.000 | 1.000 |
| Other (dummy) | 0.008 | 0.089 | 0.000 | 1.000 |

Note: Data are weighted sample means and standard deviations.

**Table 2.** Sample pairwise correlations for main variables

| | Labor Prod. | Any assoc. | Business assoc. | Non-bus assoc. | Number of customers | Credit from suppliers | Credit to customers | Loan from bank | Loan from friends | Founder has univ. ed. |
|---|---|---|---|---|---|---|---|---|---|---|
| Labor Productivity (log) | 1.000 | | | | | | | | | |
| Any association (dummy) | 0.113*** | 1.000 | | | | | | | | |
| Business association (dummy) | -0.034 | 0.164*** | 1.000 | | | | | | | |
| Non-business association (dummy) | 0.118*** | 0.980*** | 0.069*** | 1.000 | | | | | | |
| Number of customers | 0.174*** | 0.078*** | 0.026 | 0.069*** | 1.000 | | | | | |
| Obtains credit from suppliers (dummy) | 0.046* | 0.051** | 0.103*** | 0.043* | -0.021 | 1.000 | | | | |
| Provides credit to customers (dummy) | 0.081*** | 0.055** | 0.033 | 0.047** | 0.088*** | 0.192*** | 1.000 | | | |
| Loan from bank (dummy) | 0.066*** | 0.029 | 0.172*** | 0.013 | 0.047* | 0.082*** | 0.038* | 1.000 | | |
| Loan from friends/family (dummy) | -0.029 | 0.052** | 0.058*** | 0.055** | 0.038 | 0.019 | -0.011 | 0.039* | 1.000 | |
| Founder has university education (dummy) | 0.144*** | 0.093*** | 0.092*** | 0.077*** | 0.045* | 0.045** | -0.023 | 0.210*** | -0.003 | 1.000 |
| Founder has vocational education (dummy) | 0.046* | 0.029 | 0.079*** | 0.016 | 0.011 | 0.065*** | -0.045** | 0.057*** | 0.033 | -0.077*** |
| Founder has secondary education (dummy) | 0.092*** | -0.001 | -0.055** | 0.012 | 0.057** | 0.016 | 0.043** | -0.079*** | -0.040* | -0.300*** |
| Firm Age (log) | 0.027 | 0.086*** | 0.057** | 0.074*** | -0.021 | 0.046** | -0.023 | -0.014 | -0.014 | -0.057** |
| Firm is in urban area of district (dummy) | 0.209*** | 0.117*** | -0.001 | 0.117*** | 0.075*** | 0.067*** | 0.034* | 0.065*** | 0.011 | 0.123*** |
| Age of entrepreneur (log) | -0.044* | -0.000 | 0.021 | -0.005 | -0.109*** | 0.017 | -0.058*** | 0.027 | -0.024 | 0.012 |
| Has Bank Account (dummy) | 0.048** | 0.06*** | -0.003 | 0.063*** | 0.045* | 0.008 | -0.046** | 0.067*** | -0.001 | 0.218*** |
| Population density (log) | 0.192*** | -0.018 | -0.024 | -0.016 | 0.117*** | -0.001 | 0.143*** | 0.045** | -0.024 | 0.054** |
| Illiteracy rate (percent of population) | -0.231*** | 0.120*** | 0.075*** | 0.125*** | -0.064*** | 0.009 | -0.171*** | -0.022 | 0.003 | -0.020 |
| Retail Trade (dummy) | 0.027 | 0.009 | -0.099*** | 0.020 | 0.173*** | -0.010*** | 0.003 | -0.142*** | -0.008 | -0.117*** |
| Manufacturing (dummy) | -0.031 | -0.072*** | -0.009 | -0.070*** | -0.112*** | 0.036 | -0.009 | 0.000 | -0.001 | -0.040* |
| Other Services (dummy) | -0.024 | 0.017 | 0.036* | 0.012 | -0.035 | 0.071*** | 0.025 | 0.025 | 0.019 | 0.090** |
| Other (dummy) | 0.000 | 0.043** | 0.117*** | 0.030 | -0.106*** | 0.070*** | -0.005 | 0.166*** | 0.006 | 0.139*** |

***, **, * Sign at 1%, 5% and 10% levels

**Table 2.** Sample pairwise correlations for main variables (continued)

|  | Founder has voc. ed. | Founder has sec. ed. | Age of entre. | Has Bank Account | Population density | Illiteracy rate | Retail Trade | Manufact. | Other Services | Other |
|---|---|---|---|---|---|---|---|---|---|---|
| Founder has vocational education (dummy) | 1.000 | | | | | | | | | |
| Founder has secondary education (dummy) | -0.237*** | 1.000 | | | | | | | | |
| Firm Age (log) | 0.023 | -0.078*** | | | | | | | | |
| Firm is in urban area of district (dummy) | 0.073*** | 0.091*** | | | | | | | | |
| Age of entrepreneur (log) | 0.0659*** | -0.1277*** | 1.0000 | | | | | | | |
| Has Bank Account (dummy) | 0.0986*** | -0.0232 | 0.0388* | 1.0000 | | | | | | |
| Population density (log) | 0.0344 | -0.0044 | -0.0162 | -0.0224 | 1.0000 | | | | | |
| Illiteracy rate (percent of population) | -0.0144 | -0.0224 | 0.0244 | 0.0654*** | -0.7342*** | 1.0000 | | | | |
| Retail Trade (dummy) | -0.1107*** | 0.102*** | -0.0593*** | -0.0531** | -0.0333 | 0.0262 | 1.0000 | | | |
| Manufacturing (dummy) | -0.0103 | -0.0429** | 0.0589*** | 0.0067 | -0.0418** | -0.0107 | -0.5491*** | 1.0000 | | |
| Other Services (dummy) | -0.0287 | -0.0153 | 0.0368* | 0.0224 | 0.0004 | -0.0014 | -0.189*** | -0.0388* | 1.0000 | |
| Other (dummy) | 0.1534*** | -0.082*** | 0.0121 | 0.0531*** | 0.078*** | -0.0266 | -0.7051*** | -0.1447*** | -0.0498** | 1.0000 |

**Table 3.** Social Capital, Financing and Customer Base Mediators, and Labor Productivity

|  | (1) | (2) | (3) | (4) | (5) | (6) |
|---|---|---|---|---|---|---|
| **Estimation Method** | Probit | Probit | Probit | Probit | Interval | OLS |
| **Dependent Variable** | Provides credit to customers | Gets credit from suppliers | Loan from friends/family | Loan from bank | Number of customers (log) | Labor Productivity |
| **Social Capital** | | | | | | |
| Any association | 0.196*** | 0.203** | 0.278*** | -0.050 | 0.155** | 0.350*** |
|  | (3.11) | (2.01) | (2.81) | (-0.37) | (2.20) | (4.22) |
| **Firm Characteristics** | | | | | | |
| Owner has university | -0.188 | 0.239 | -0.146 | 0.859*** | 0.372*** | 1.059*** |
|  | (-1.54) | (1.36) | (-0.83) | (4.54) | (2.84) | (6.58) |
| Owner has vocational | -0.223 | 0.435** | 0.119 | 0.346 | 0.268* | 0.635*** |
|  | (-1.53) | (2.32) | (0.64) | (1.43) | (1.75) | (3.36) |
| Owner has secondary | 0.042 | 0.237** | -0.239** | -0.013 | 0.198*** | 0.475*** |
|  | (0.63) | (2.19) | (-2.38) | (-0.08) | (2.65) | (5.43) |
| Firm Age (log) | 0.003 | 0.091* | -0.034 | -0.006 | 0.069* | 0.154*** |
|  | (0.08) | (1.81) | (-0.68) | (-0.09) | (1.85) | (3.50) |
| Firm is in urban area | -0.026 | 0.255*** | 0.088 | 0.307** | -0.062 | 0.497*** |
|  | (-0.39) | (2.58) | (0.88) | (2.16) | (-0.83) | (5.58) |
| Age of entrepreneur (log) | -0.236** | -0.078 | -0.196 | 0.014 | -0.538*** | -0.363** |
|  | (-2.12) | (-0.45) | (-1.15) | (0.06) | (-4.33) | (-2.48) |
| Has Bank Account | -0.155 | -0.085 | 0.039 | 0.076 | 0.393*** | 0.037 |
|  | (-1.10) | (-0.44) | (0.20) | (0.35) | (2.69) | (0.21) |
| **District Characteristics** | | | | | | |
| Population density | 0.025 | -0.040 | -0.076** | 0.021 | 0.089*** | -0.052* |
|  | (1.10) | (-1.18) | (-2.16) | (0.47) | (3.52) | (-1.68) |
| Illiteracy rate | -0.016*** | -0.002 | -0.007 | 0.002 | -0.001 | -0.035*** |
|  | (-4.59) | (-0.42) | (-1.41) | (0.28) | (-0.18) | (-7.99) |
| Constant | 0.634 | -1.642** | -0.359 | -2.606*** | 5.017*** | 16.306*** |
|  | (1.45) | (-2.42) | (-0.54) | (-2.80) | (10.34) | (28.31) |
| **Sector (Industry) Dummies** | Yes | Yes | Yes | Yes | Yes | Yes |
| **Observations** | 1,954 | 1,844 | 1,971 | 1,971 | 1,539 | 1,609 |
| **R-Squared** | 0.0363 | 0.0477 | 0.0227 | 0.154 | --- | 0.150 |

Note: This table presents the reduced form regressions. Columns 1-4 report estimates from a probit model. In column 1, the dependent variable is *Provides credit to customers,* an indicator variable that is equal to one if the owner provides credit to customers and zero otherwise. In column 2, the dependent variable is *Gets credit from suppliers*, which is equal to one when the owners get credit from suppliers and zero otherwise. In column 3 the dependent variable is *Loan from friends/family*, which is equal to one when the owners get credit from family and friends and zero otherwise. In column 4 the dependent variable is *Loan from bank*, which is equal to one when the owners get loans from banks and zero otherwise. Column 5 reports coefficients of interval regression. The dependent variable is the *Number of customers (log)*, the natural logarithm of the number of customers. In the survey, the number of customers is reported as a range. As a result, we estimate the model as an interval regression, which is generalization of the Tobit model (Stata 2007). We use the command 'intreg' in Stata, which is described in detail in the Stata base reference manual. Column 6 reports the coefficient of OLS regression where the dependent variable is *Labor Productivity*, the natural log of ratio of sales to the number of workers. *Any association,* coded one if the owner belongs to any association and 0 otherwise. *Owner has university education (dummy)* is one if the owners have a university education and zero otherwise. *Owner has vocational education (dummy)* is one if the owner has vocation education and zero otherwise. *Owner has secondary education (dummy)* is one if the owner has secondary education and zero otherwise. Firm Age (log) is the natural logarithm of the firm's age. *Firm is in urban area (dummy)* is one if the firm is in an urban area and zero otherwise. *Age of entrepreneur (log)* is the natural logarithm of the age of the entrepreneur. *Has Bank Account* is one if the entrepreneur has a bank account. *Population density* is measured as the

**Table 4.** Summary of Direct and Indirect Effects

| | |
|---|---:|
| Direct effect | |
|   Social capital → labor productivity | 0.271 |
| Indirect effect | |
|   Social capital → credit from suppliers → credit to customers → number of customers → labor productivity | 0.034 |
|   Social capital → credit to customers → number of customers → labor productivity | 0.013 |
|   Social capital → credit from suppliers → number of customers → labor productivity | -0.013 |
|   Social capital → loans from bank → number of customers → labor productivity | -0.004 |
|   Social capital → loans from friends & family → number of customers → labor productivity | 0.027 |
| Total Indirect Effect (Σ |Indirect effects |)[a,b] | |
|   Social capital → labor productivity | 0.091 |
| Total Effect (Direct + indirect)[c] | |
|   Social capital → labor productivity | 0.362 |
| Percent mediated by indirect effects[d] | 25.14% |

Note: Results based on SEM model from Figure 1. respectively. [a] We follow Alwin and Hauser (1975) and the literature on "inconsistent" mediation MacKinnon et al. (2007), which suggests researchers use the absolute value of effect sizes when calculating the proportion mediated when some paths have opposite signs. [b] 0.091 = 0.034 + 0.013 + |-0.013| + |-0.004| + 0.027. [c] 0.362 = 0.271 + 0.091. [d] 25.14% = 0.091/0.362. The ***, **, * denote significances at the 1%, 5%, and 10% levels.